\documentclass{aastex}          
\usepackage{spr-astr-addons} 
\usepackage{epsfig}
\usepackage{url}

\begin{document}
%
\title{Reexamining generalized Chaplygin gas with the sign-changeable interaction}


\author{Ping Xi} \and \author{Ping Li} \affil{Shanghai United Center for
Astrophysics(SUCA), Shanghai Normal University, 100 Guilin Road,
Shanghai 200234,China} \email{xiping@shnu.edu.cn}


\begin{abstract}
In this paper, we reexamine the generalized Chaplygin gas (GCG) cosmology with the sign-changeable interaction. The dynamical analysis show that there exists de-Sitter attractors in this model, which means that the late-time behaviors of the model is insensitive to the initial condition and thus alleviates the coincidence problem. Furthermore, we find that this interaction oscillates and tends to zero finally by numerical calculations of the nonlinear equations. In some specific cases of initial conditions, the interaction is positive (the energy transition from dark energy to dark matter) at high redshift while it is negative (the energy transition from dark matter to dark energy) at low redshift for suitable ranges of the parameters.
\end{abstract}


%
\section{Introduction}
The current observations, such as SNeIa (Supernovae typeIa), CMB (Cosmic Microwave Background) and large scale structure,converge on the fact that dark energy (DE), a spatially homogeneous and gravitationally repulsive energy component, accounts for about $68$ \% of the total energy density of universe and dark matter (DM) accounts for $27$ \%. Some heuristic models that roughly describe the observable consequences of dark energy were proposed in recent years, such as the cosmological constant, quintessence \citep{pee03,li02}, phantom \citep{cal02,liu03} and generalized Chaplygin gas (GCG) \citep{mak03,zha06} stemming form the Chaplygin gas. And GCG has a very simple equation of state
\begin{equation}
p_g=-A/\rho_g^\alpha.
\end{equation}
where $p_g$ and $\rho_g$ are the pressure and the density of GCG, $A$ is the parameter and $\alpha$ is in the range of ($0, 1$). Latest observations \citep{ade15} indicate the possibility of dark energy with $w<-1$. If this is the case, the Universe suffers from various catastrophic fates including big rip \citep{cal03,ste05}, little rip \citep{fra12} and other future singularity \citep{mci02}.We found that the phantom-like GCG \citep{hao05} can avoid the catastrophic fate of phantom. But GCG also encounters a basic problem in almost any cosmological model,that is, why the values of the energy densities of dark energy and dark matter are of the same order at present. It is called the coincide problem.

On the other hand, the interaction between DE and DM is studied widely. One of the important motivations is to alleviate the cosmological coincide problem. Due to lacking a guidance from fundamental physics, this interaction is discussed only on phenomenological level which is usually assumed as in the form of $3\beta H\rho_m$, $3\beta H\rho_{de}$ or $3\beta H(\rho_m+\rho_{de})$ ($\rho_m$ and $\rho_{de}$ are the densities of DM and DE, $H$ is the Hubble parameter and $\beta$ is a dimensionless quantity.). Various observation data are used to investigate this interaction. The CMB+BAO+SN+$H_0$ data show that the interaction appears an oscillating aspect and passes through the noninteracting at $0.45\lesssim z\lesssim0.9$ \citep{cai10} which derives a modified form of the interaction including the decelerating parameter $q$ \citep{wei11}. The current Planck+SN/RSD suggests the late-time interaction commences at $z\sim0.9$ \citep{sal14}. And another observations such as the Sandage-Loeb test \citep{gen15} or BOSS \citep{abd14} indicate that there may be this interaction at high redshift and DE decays into DM.

Recently, the interacting modified Chaplygin gas models are investigated by Xu et al. \citep{xu12,xu13}. To analyze the dynamical system, the authors employed the dimensionless variables as follows
\begin{eqnarray}
x&=&\frac{\kappa^2\rho_g}{3H^2},\\
y&=&\frac{\kappa^2p_g}{3H^2},\\
N&=&\ln{a}.
\end{eqnarray}
Unfortunately, their autonomous system has a singular point at $x=0$ under this choice, so that it does not satisfy the existence and uniqueness theorem of differential equations in whole phase space. To avoid this singularity, we choose new dimensionless variables (see section 2) and reexamine the dynamics of the GCG cosmology with the sign-changeable interaction. Using the dynamical system techniques, we find that only the interacting GCG cosmology with $Q=3\beta qH\rho_m$ is physically acceptable and there is a pair of negative and positive attractors in this model. Thus, the late-time behaviors of this model is insensitive to the initial condition and alleviates the coincidence problem. Via numerical calculations, we find that this interaction oscillates and tends to zero finally. In some specific cases of initial conditions, the interaction is positive at high redshift while it is negative at low redshift for suitable ranges of the parameters.

\section{Nonlinear autonomous system}
We assume that there are two dark fluids in our universe: GCG ($w_g=p_g/\rho_g$)and DM ($w_m=0$). For this interacting GCG cosmological dynamical system, the corresponding Einstein and motion equations could be written as
\begin{eqnarray}
H^2&=&\frac{\kappa^2}{3}(\rho_g+\rho_m)\label{FEQ},\\
\dot{H}&=&-\frac{\kappa^2}{2}(\rho_g+\rho_m+p_g),\\
\dot{\rho_m}&=&-3H\rho_m+Q,\\
\dot{\rho_g}&=&-3H(\rho_g+p_g)-Q,
\end{eqnarray}
where $\kappa^2=8\pi G$ and $Q$ is the interactions with the forms $3\beta qH\rho_i$ ($i=1,2,3$, $\rho_{1,2,3}=\rho_g+\rho_m$, $\rho_g$ or $\rho_m$). The over dot denotes a derivative with respect to cosmic time $t$.

To analyze the dynamical system, we introduce the following dimensionless variables
\begin{eqnarray}
x&=&\frac{3H^2}{\kappa^2\rho_g},\\
y&=&\frac{\kappa^2p_g}{3H^2},\\
N&=&\ln{\frac{a}{a_0}},
\end{eqnarray}
where $a$ is the scale factor and the subscript zero represents the quantity given at the present epoch (Here, we fix $a_0=1$.). Eq. (9) is a different variable from Eq. (7) in \citep{xu12}. Then, we obtain the equations of the corresponding autonomous system as follows
\begin{eqnarray}
\frac{dx}{dN}=3x[1+xy+\frac{\beta}{2}x(1+3y)\frac{\kappa^2\rho_i}{3H^2}]-3x(1+y),\\
\frac{dy}{dN}=3\alpha y[1+xy+\frac{\beta}{2}x(1+3y)\frac{\kappa^2\rho_i}{3H^2}]+3y(1+y).
\end{eqnarray}
Obviously, this system satisfies the fundamental theorem of differential equations in $x-y$ plane. On the contrary, the autonomous system has a singular point in previous work \citep{xu12}, so our autonomous system is an important advantage. Moreover, we obtain the following expression by Friedmann equation (\ref{FEQ}),
\begin{equation}
\Omega_g+\Omega_m=1,
\end{equation}
where $\Omega_g\equiv1/x$ and $\Omega_m\equiv\frac{\kappa^2\rho_m}{3H^2}$ are the density parameters for GCG and DM, respectively. So, the physically significant value of $x$ is in $[1, +\infty)$ which corresponds to the value of $\Omega_g$ in [$0,1$].

The critical points of the autonomous system ($x_c, y_c$) are the solutions to the equations of $dx/dN=0$ and $dy/dN=0$. We find that there is only one physical critical point for the dynamical system with $Q=3\beta qH(\rho_{g}+\rho_{m})$ or $Q=3\beta qH\rho_{g}$ (see the Appendix) and there are two physical critical points in the system with $Q=3\beta qH\rho_m$, as listed in Table 1. Therefore, only the interacting GCG cosmology with $3\beta qH\rho_m$ is physically acceptable and the other two including that in \citep{xu12} aren't acceptable on basis of the succession of cosmology epochs. In the next section, we will dynamically analyze the GCG cosmology with $Q=3\beta qH\rho_m$ in detail.
\begin{table}[h]
\small
\caption{\label{table:cp-3} All the critical points and their physical conditions for the interacting GCG cosmology with $Q=3\beta qH\rho_m$. }
  \begin{tabular}{c|c|c|c}
  \hline
  \hline
   Critical Points & $x_{c}$ & $y_{c}$ & Physical Condition	\\
  \hline
  $P_{1}$               &$0$                              & $0$ 		 & always unphysical\\
  \hline
  $P_{2}$               &$0$              & $\frac{2+\alpha(2-\beta)}{3\alpha\beta-2}$ 		 & always unphysical\\
  \hline
  $P_{3}$               &$1$                              & $0$  & $0<\alpha\leq1$\\
  \hline
  $p_{4}$               &$1$                              & $-1$ & $0<\alpha\leq1$\\
  \hline
  \hline
  \end{tabular}
 \end{table}

\section{The numerical results of the interacting GCG cosmology with $Q=3\beta qH\rho_m$}
\subsection{Dynamical behaviors of the autonomous system}
To investigate the property of the physical critical points ($x_c$, $y_c$) in Table 1, we write the variables near the two critical points in the form $x=x_c+\delta x$ and $y=y_c+\delta y$ with the perturbations $\delta x$ and $\delta y$. Substituting the expression into the autonomous system of equations (12)-(13), we can obtain the corresponding eigenvalues of critical points $p_3$ and $p_4$:
\begin{eqnarray}
&p_3&: (3(1+\alpha),\frac{3}{2}\beta),\nonumber\\
&p_4&: (-3(1+\alpha),-3(1+\beta)).
\end{eqnarray}
The properties of the critical points are shown in Table 2. Although the property of critical points are obtained from the linearized system of the nonlinear system near the critical point in the Table 2, the linearization theorem shows that their properties are preserved for the nonlinear system (12)-(13) because the critical points of the nonlinear system are hyperbolic. The property of the critical point changes with the parameters $\alpha$ and $\beta$. By analysis of bifurcations for this system \citep{fen12}, we can divide in the $\alpha-\beta$ plane into 5 domains shown in Fig. 1. In each domain, all the critical points have same type. We find that critical point $p_4$ is a late-time de Sitter attractor in the case of $\beta>-1$, which means that this interacting GCG cosmology is an elegant scheme.
\begin{figure}
\includegraphics[height=2.1in,width=2.5in]{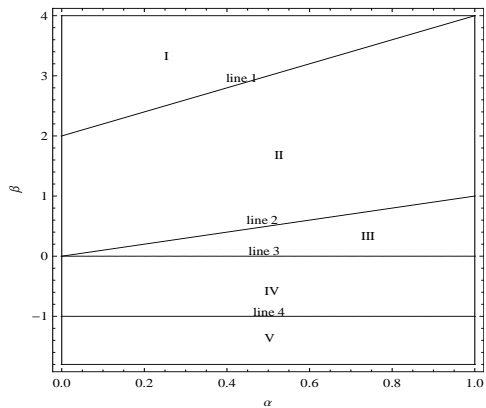}
\caption{The two-parameter plane is divided into 5 domains according to the type of critical point. All the critical points have same property in each domain.}
\end{figure}
\begin{table*}
\begin{center}
\small
\caption{\label{table:cpp} The physical properties of critical points of the autonomous system (12)-(13) with $Q=3\beta qH\rho_m$.}
 \begin{tabular}{c|c|c|c|c}
  \hline
  \hline
  Domain    & $P_{3}$            & $P_{4}$            	\\
  \hline
  \hline
  I               & source                      & sink\\
  \hline
  line 1          & spiral source or source     & sink\\
  \hline
  II              & source                      & sink \\
  \hline
  line 2          & source                      & sink or spiral sink\\
  \hline
  III             & source                        & sink		\\
  \hline
  line 3          & linear theory fails to determine its property       & sink\\
  \hline
  IV             & saddle                      & sink  \\
  \hline
  line 4         & saddle                      & linear theory fails to determine its property\\
  \hline
  V              & saddle   	     & saddle \\
   \hline
   \hline
  \end{tabular}
  \end{center}
  \end{table*}

Then, we discuss the dynamical behaviors of the interacting GCG cosmology numerically. The evolution trajectories of this model with different initial conditions are shown in Fig.2. One can see that the Universe experiences the transition from a matter dominated era to dark energy dominated era which is insensitive to the initial condition. The corresponding critical points are $(1,0)$ and $(1,-1)$, respectively. GCG can behave as quintessence or phantom between these two phases of the Universe. Especially, phantom-like GCG in this model won't lead to big rip for there existing a de Sitter attractor \citep{hao05}. On the other hand, GCG cosmology is a special case of the interacting GCG cosmology with $\beta=0$. The corresponding evolution curves are plotted in Fig.3, which are different from those in the interacting GCG cosmology.
\begin{figure}
\includegraphics[height=2.1in,width=2.5in]{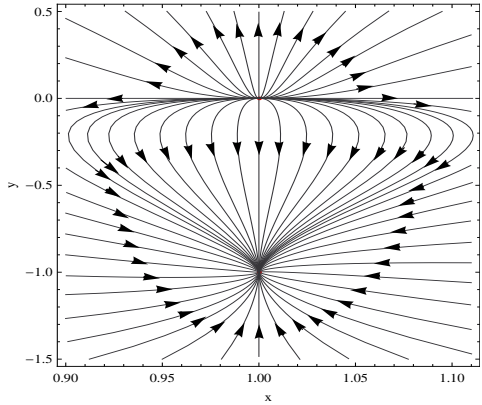}
\caption{The phase diagrams of ($x$,$y$) for the interacting GCG cosmology with $\beta>-1$. The heteroclinic orbit connects the critical points $p_3$ to $p_4$. We take $\alpha=0.4$ and $\beta=1.2$.}
\end{figure}
\begin{figure}
\includegraphics[height=2.1in,width=2.5in]{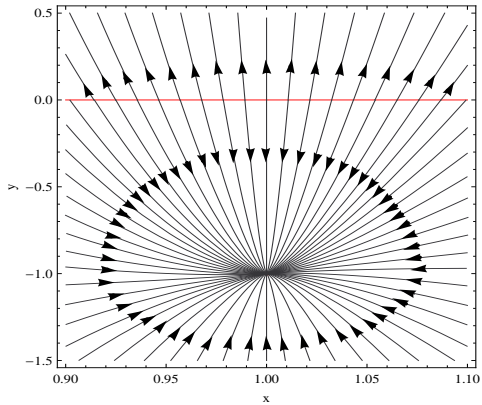}
\caption{The phase diagrams of ($x$,$y$) for the interacting GCG cosmology with $\beta=0$. We take $\alpha=0.4$.}
\end{figure}

\subsection{The oscillating interaction}
The interaction $Q=3\beta qH\rho_m$ in this model plays an important role of the transformation between GCG and DM. It is interesting to study the evolution of the interaction. Via numerical evolution of the nonlinear equations (12)-(13), we find that this model contributes an oscillating aspect to the interaction. This result is supported by those in Ref. \citep{cai10}. And the oscillating interaction tends to zero eventually which means that there is no DM at late-time in this cosmology model, as shown in Fig. 4-5. In these two figures, we can see that $Q$ increases firstly starting at $Q=0$, then decreases from $Q_{max}$ down to $Q_{min}$ (crossing $Q=0$), and increases from $Q_{min}$ to zero. If $Q$ is positive, it indicates that the decay direction is from GCG to dark matter; while $Q$ is negative, the transition direction is from dark matter to GCG; if $Q=0$, it means that there does not exist the interaction between GCG and DM. Then, we will discuss the relation between $Q$ and the parameters $\alpha$ and $\beta$ for given initial conditions. In Fig. 4, we choose $\alpha=0.4$, $x(0)=1.01$ and $y(0)=-0.7$ to study the evolution of $Q$ for different $\beta$. These interactions at $z=0.9$ are negative which is similar to that of Ref. \cite{sal14}. While those at $z=2.34$ are positive, which is alike to that of Ref. \citep{abd14}. As $\beta$ increases, the interaction at $z=0.9$ decreases while that at $z=2.34$ increases. In Fig. 5, fixing $x(0)=1.01$, $y(0)=-0.7399$ and the negative values of the interactions at $z=0.9$ being equal, we discuss the evolution of the interaction for different $\alpha$ and $\beta$. Obviously, the interactions at $z=2.34$ increases as $\alpha$ and $\beta$ decrease. And we also consider the parameters $\alpha$ and $\beta$ for the oscillating interaction with $Q(z=0.9)<0$ and $Q(z=2.34)>0$. In fig. 6, fixing $x(0)=1.01$ and $y(0)\in[-0.7399, -0.6535]$, the range of the parameters $\alpha$ and $\beta$ are obtained by numerical computing. In fig. 7, choosing $x(0)=1.42$ and $y(0)\in[-0.7399, -0.6535]$, the range of $\alpha$ and $\beta$ for the interaction with $Q(z=0.9)>0$ are calculated.
\begin{figure}
\includegraphics[height=2.1in,width=2.5in]{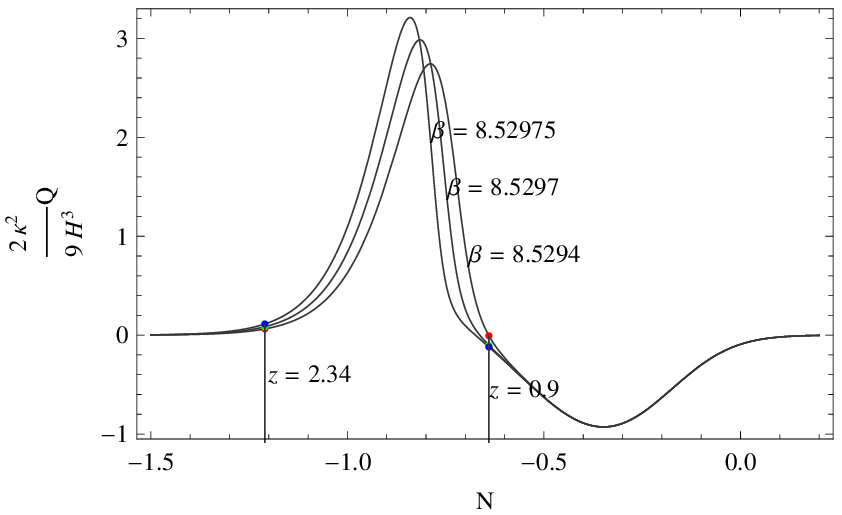}
\caption{The evolution of $Q$ are shown with different $\beta$.}
\end{figure}
\begin{figure}
\includegraphics[height=2.1in,width=2.5in]{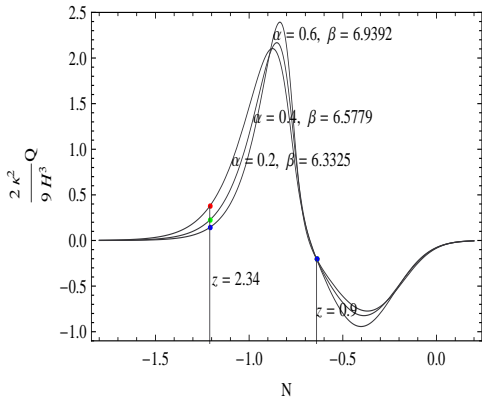}
\caption{The evolution of $Q$ are shown for different $\alpha$ and $\beta$.}
\end{figure}
\begin{figure}
\includegraphics[height=2.1in,width=2.5in]{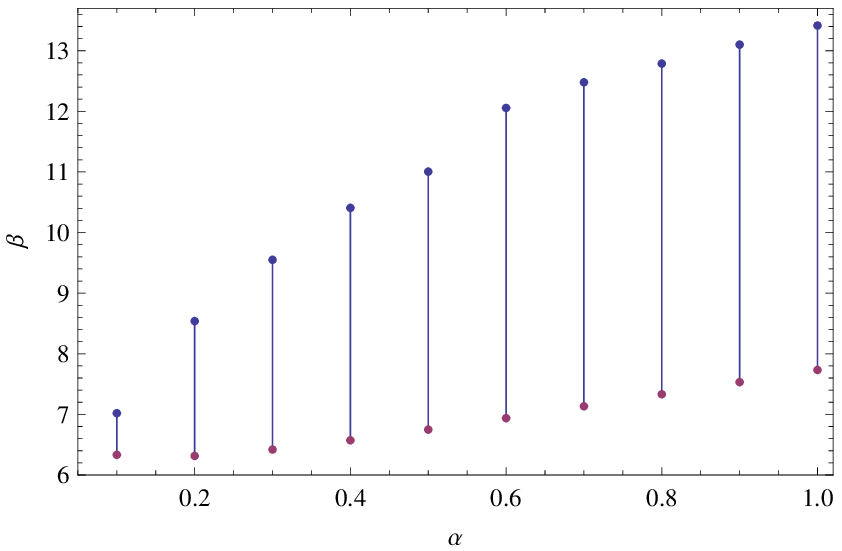}
\caption{The parameters $\alpha$ and $\beta$ for $Q(z=2.34)>0$ and $Q(z=0.9)<0$ are shown.}
\end{figure}
\begin{figure}
\includegraphics[height=2.1in,width=2.5in]{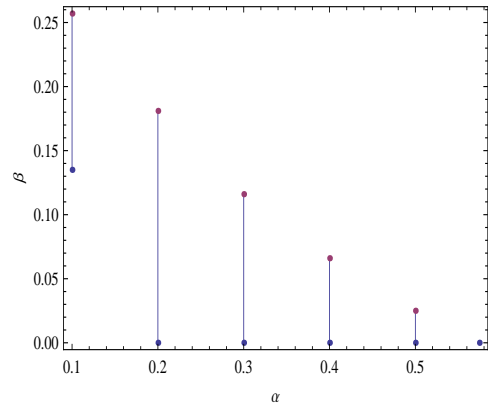}
\caption{The parameters $\alpha$ and $\beta$ for $Q(z=0.9)<0$ are shown.}
\end{figure}
\section{Conclusion}
In this paper, we reexamine the dynamics of the GCG cosmology with the sign-changeable interaction. Introducing a new dynamical variable, we obtain the dynamical system for this model, which is different from those in \citep{xu12}. In the light of the series of the dominated eras in cosmology, only the GCG cosmology with $3\beta qH\rho_m$ is physically acceptable. It means that what is discussed in \citep{xu12} isn't physically acceptable. Using the dynamical system techniques to study this model, the results are obtained as follows:
(i)There exists de-Sitter attractors in the late-time of this model, which alleviates the coincidence problem.
(ii)This interaction has an oscillation form and will tend to zero eventually. In some specific cases of initial conditions, the interaction is positive at high redshift while it is negative at low redshift for suitable ranges of the parameters.

\acknowledgments {We thank Prof. X. Z. Li for helpful discussions. This work is supported by National Science Foundation of China grant. No. 11205102 and Innovation Program of Shanghai Municipal Education Commission (12YZ089).}



 \appendix{Appendix}
\begin{table}[h]
\centering
\caption{\label{table:cp-1} All the critical points and their physical conditions for the interacting GCG cosmology with $Q=3\beta qH(\rho_g+\rho_m)$. }
  \begin{tabular}{c|c|c|c}
  \hline
  \hline
   Critical Points & $x_{c}$ & $y_{c}$ & Physical Condition	\\
  \hline
  $P_{1}$               &$0$                              & $0$ 		 & always unphysical\\
  \hline
  $P_{2}$               &$0$              & $-1-\alpha$ 		 & always unphysical\\
  \hline
  $P_{3}$               &$\frac{1}{1+\beta}$                              & $-1$  & $-1<\beta\leq0$\\
  \hline
  \hline
  \end{tabular}

\end{table}
\begin{table}[h]
\centering
\caption{\label{table:cp-2} All the critical points and their physical conditions for the interacting GCG cosmology with $Q=3\beta qH\rho_g$. }
  \begin{tabular}{c|c|c|c}
  \hline
  \hline
   Critical Points & $x_{c}$ & $y_{c}$ & Physical Condition	\\
  \hline
  $P_{1}$               &$0$                              & $0$ 		 & always unphysical\\
  \hline
  $P_{2}$               &$0$              & $-\frac{2+\alpha(2+\beta)}{2+3\alpha\beta}$ 		 & always unphysical\\
  \hline
  $P_{3}$               &$1-\beta$                              & $-1$  & $0\leq\beta\leq1$\\
  \hline
  \hline
  \end{tabular}

\end{table}

\end{document}